\documentclass[twocolumn,showpacs,superscriptaddress]{revtex4}
\usepackage{amssymb,amssymb}
\usepackage[pdftex]{graphicx}

\def\Vec#1{\mbox{\boldmath $#1$}}

\begin{document}
\title{Dynamics of Polymer Decompression: Expansion, Unfolding and Ejection}
\author{Takahiro Sakaue}
\email{sakaue@scphys.kyoto-u.ac.jp}
\affiliation{Fukui Institute for Fundamental Chemistry, Kyoto University, Kyoto 606-8103, Japan}
\author{Natsuhiko Yoshinaga}
\affiliation{Department of Physics, The
University of Tokyo, Tokyo 113-0033, Japan}

\begin{abstract}
 The dynamics of polymer decompression from a compact state to swollen
 conformation can be formally described as nonlinear diffusion. We
 discuss two basic examples: (i) the expansion, or unfolding from a
 compact state, and (ii) the ejection of a compressed polymer through a
 pore. The problem can be solved exactly for case (i), but not for
 case (ii). Even in such situations, a scheme called uniform
 approximation is shown to be useful to get a physical insight
 involved. Its application to the case (ii) is able to account for
 conflicting numerical data in a consistent way.
\end{abstract}


\maketitle

Macromolecules assume compact conformations in certain
situations. Examples include DNA in living cells, proteins in native
states and  other polymers in poor solvent conditions or under a
compression field.
Once released from the condition,
these polymers expand to swollen coiled state, which is characterized by developed fluctuations.
This expansion process is of interest in two different
contexts. First, this process is relevant to coil-globule transitions,
thus, regarded as a fundamental topic in polymer
science~\cite{grosberg,Xu}. 
Unfortunately, there are comparatively fewer past studies on this
expansion process than on the folding (coil to globule) process.~\cite{deGenne-kinetics, Halperin, Yoshinaga}
The second case of interest is encountered in the field of confined polymers~\cite{Kasianowicz}.
A recent advance in nanoscale fabrications and single-chain experiments allows one to manipulate and observe individual polymers, thereby facilitating a number of potential applications in biological as well as nanoscale sciences~\cite{aHL,DrivenDNA,Hoagland,Craighead,DNAinNonochannels,Ichikawa}.
The aim of the present Letter is to provide a unified framework to describe decompression processes engaged in such situations.

\begin{figure}[t]
\includegraphics[width=0.28\textwidth]{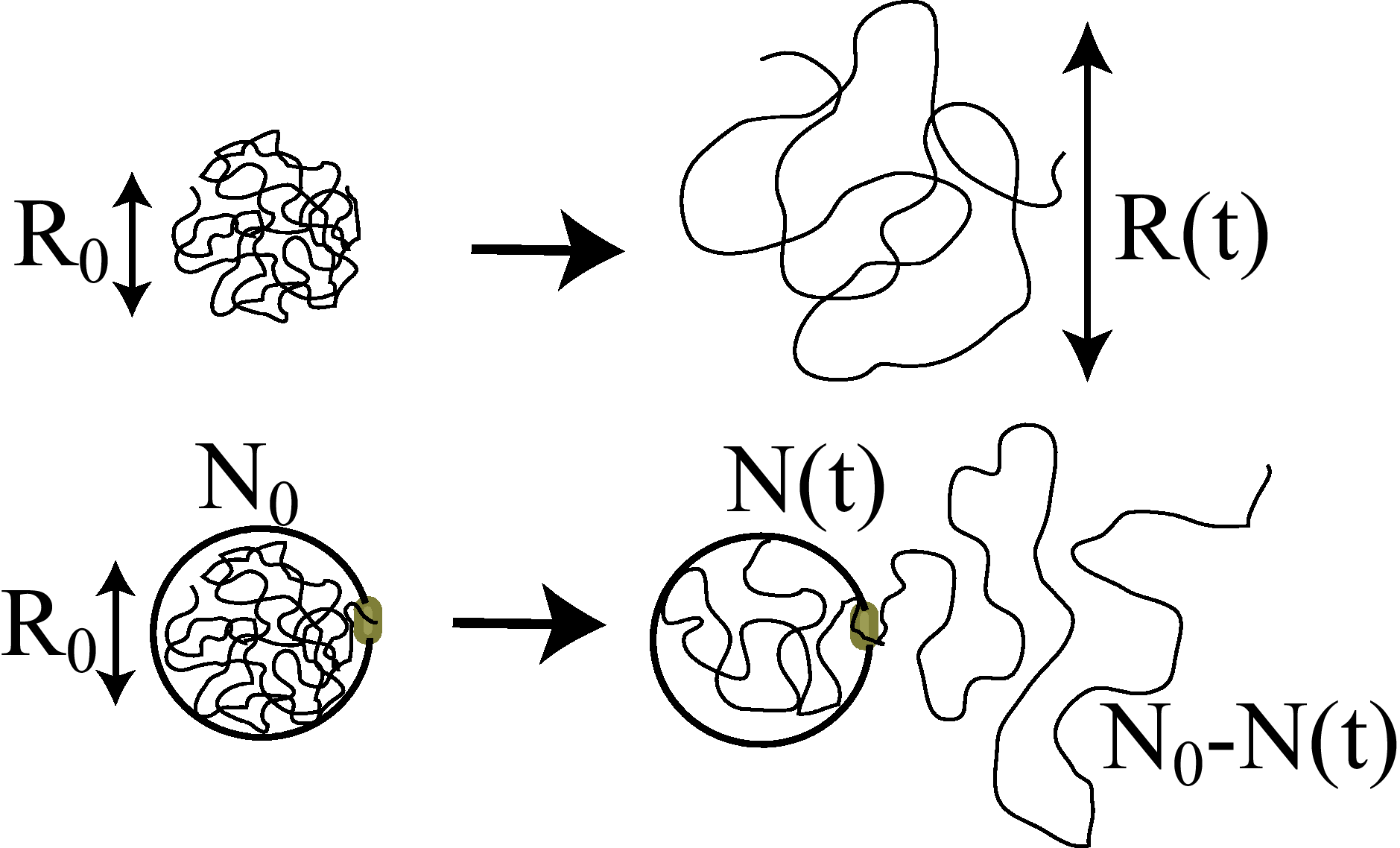}
\caption{Schematics of [top] expansion (unfolding) and [bottom] ejection processes.}
\label{fig1}
\end{figure}

Naively, the problem is analogous to the diffusion of molecular gases, which are initially confined in a finite-size box. 
For our case, the essential difference from this simple example lies in the {\it connectivity} of monomeric units into a string, in which a small entropy does not play any major role.
By taking such polymeric natures correctly into account, it is naturally formulated as a nonlinear diffusion problem, which serves as a basis for various decompression processes such as (i) the expansion, i.e., unfolding, and (ii) the ejection of a geometrically compressed polymer from a narrow pore (Fig.~\ref{fig1}).

In the case of polymer expansion (case (i)), a symmetry allows for an analytic solution. However, this is a rather special case, and the nonlinearity in the partial differential equation usually requires a numerical calculation. In such a case, we propose a scheme {\it uniform approximation}, in which the nonuniformity of the segment distribution is totally neglected. It allows one to obtain an approximate solution easily, which can yet be reasonably compared with the exact solution, provided that mechanisms of the driving force and the dissipation are correctly identified.
We apply this scheme to the case of polymer ejection (case (ii)) and demonstrate a very good agreement with the numerical simulations reported so far.


Consider a compressed globular chain, the size $R_0$ of which is substantially smaller than the relaxed size $R_F=aN_0^{\nu}$ in bulk ($a$, $N_0$ are the size, the number of monomers and $\nu$ is the Flory exponent).
The corresponding average volume fraction $\phi_0 \simeq N_0 a^3/R_0^3 \ll 1$ is assumed to be small.
Here and in what follows, we assume that isotropic compression, i.e., in a sphere, but generalization to anisotropic compression is straightforward.
When given an escaping route, the polymer starts to be decompressed toward the relaxed coil.
The time evolution of volume fraction $\phi ({\Vec r})$ obeys the continuity equation;
$\partial \phi ({\Vec r})/\partial t = -{\Vec \nabla} \cdot {\Vec j}({\Vec r})$
with the flux 
${\Vec j}({\Vec r}) =  -L({\Vec r}) [{\Vec \nabla} \mu({\Vec r})] \phi({\Vec r})$.
In order to obtain the chemical potential $\mu({\Vec r}) = \partial f({\Vec r})/\partial \phi({\Vec r})$ and the transport coefficient $L({\Vec r})$, we utilize the fact that the compressed polymer resembles a semidilute solution, which amounts to say that the compressed polymer is viewed as a dense piling of blobs of size $\xi({\Vec r}) \simeq a \phi({\Vec r})^{\nu/(1-3\nu)}$. Such a situation is called a {\it strong confinement} (SC) in Refs.~\cite{Sakaue_EPL, Sakaue_Macro2} (see also eq.~(\ref{F})).
Then, the excess free energy density due to the compression and the dissipation mechanism are, respectively, evaluated by assigning $\sim k_BT$ (thermal energy) and the Stokes friction per blob~\cite{deGennes}:
$f({\Vec r}) \simeq k_BT/\xi({\Vec r})^3$
and
$L({\Vec r}) \simeq g({\Vec r})/(\eta \xi({\Vec r}))$
where $\eta$ is the solution viscosity and $g$ is the number of monomers per blob ($\xi \simeq a g^{\nu}$).
This leads to ${\Vec j} = -a^2/\tau_0 \phi({\bf r})^{\nu/(3\nu-1)} \nabla \phi({\bf r})$ with the monomer scale time constant $\tau_0 \simeq \eta a^3/(k_BT)$.
The same expression is obtained from slightly different route, which utilizes the semidilute solution analogy in a more transparent way:
${\Vec j} = -D_c({\Vec r}) {\Vec \nabla} \phi({\Vec r})$,
where
$D_c({\Vec r}) \simeq k_BT /(\eta \xi({\Vec r}))$
is the cooperative diffusion coefficient~\cite{deGennes}.
Using this flux and the continuity equation, we arrive at the following:
\begin{eqnarray}
\partial_t \phi = (a^2/\tau_0) \nabla [(\phi({\bf r}))^{m} \nabla (\phi({\bf r}))],
\label{eq_solve}
\end{eqnarray}
with $m=\nu/(3\nu-1)$ \cite{note2}.
This equation constitutes a basis for the decompression process and should be solved with appropriate initial and boundary conditions specified for each case of interest.

\paragraph{Expansion and unfolding}
There are several ways to achieve decompression. The simplest is to switch off the external compression field abruptly at $t=0$. Then, the polymer will expand isotropically toward the relaxed coil(Fig.~\ref{fig1}). In the context of the unfolding by the temperature (more generally, solvent quality) jump across the $\theta$ temperature ($T_{ini} \ \rightarrow T_{fin}$), the initial compact state can be viewed as a dense piling of thermal blobs of size $\xi_{th}= a/|\varsigma_{ini}| = a g_{th}^{1/2}$, where $\varsigma = (T-\theta)/\theta$ is the reduced temperature. Hence, the initial size of the polymer is $R_0 \simeq \xi_{th}(N_0/g_{th})^{1/3} = a N_0^{1/3}/|\varsigma_{ini}|^{1/3}$.

\begin{figure}[h]
\includegraphics[width=0.48\textwidth]{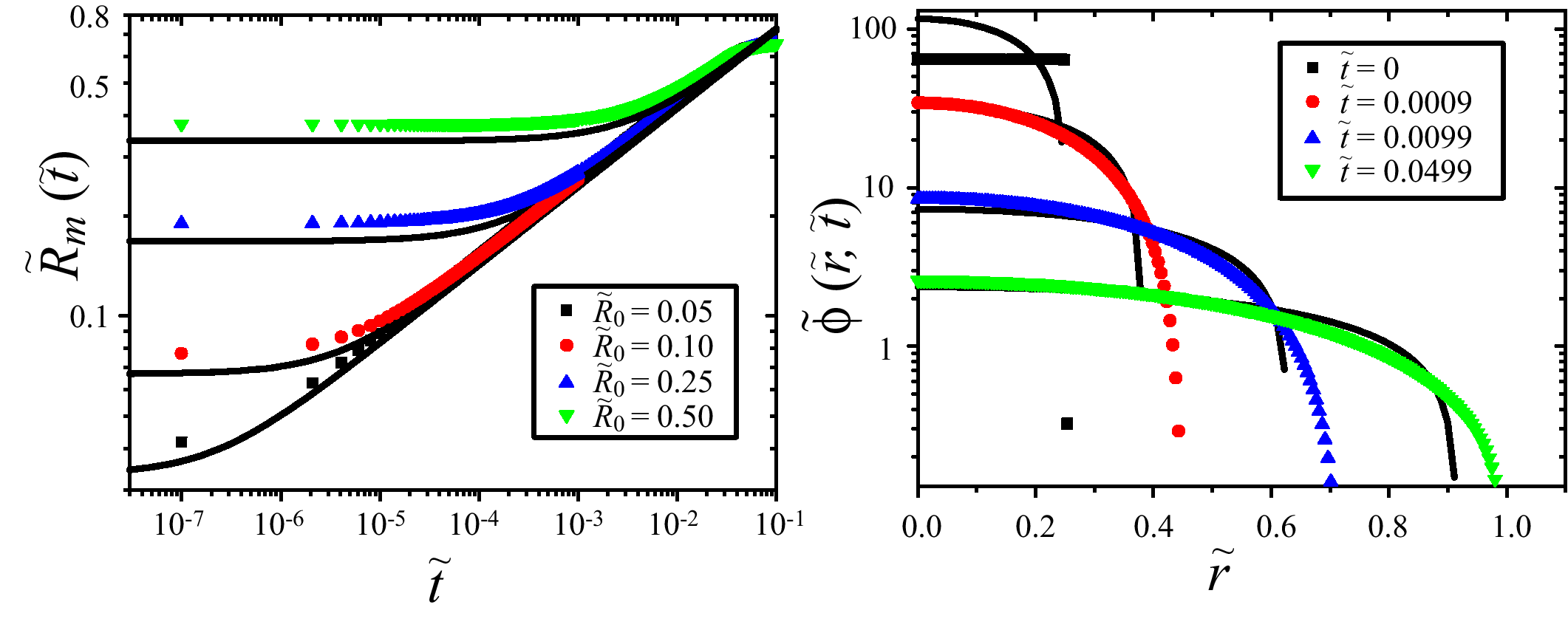}
\caption{Solutions of nonlinear diffusion equation (solid lines) (with $\nu=3/5$) together with numerical
 results for initial condition of step-like distributions (points).}
\label{fig.fp}
\end{figure}

Let us first make eq.~(\ref{eq_solve}) dimensionless by introducing
a scaled coordinate (radial component) ${\tilde r}=r/R_F$, volume fraction
${\tilde \phi}({\tilde r})=\phi(r)/\phi_F$ and time ${\tilde t}=t/\tau$,
where $\phi_F =a^3 N_0 /R_F^3$ denotes the monomer volume fraction in a
relaxed state.
By setting the time unit $\tau = \tau_{ex}= \tau_0 N_0^{3\nu}$, 
we can rewrite eq.~(\ref{eq_solve}) as
\begin{eqnarray}
\partial_{{\tilde t}} {\tilde \phi ({\tilde r})} =  {\tilde r}^{-2}  \partial_{{\tilde r}}[{\tilde r}^2  {\tilde \phi}({\tilde r})^{m}  \partial_{{\tilde r}} {\tilde \phi}({\tilde r})].
\label{eq_nondim}
\end{eqnarray}
The characteristic time for expansion $\tau_{ex}$ depends only on the
relaxed coil size $R_F$ and not on the initial size.
Therefore, it has the same scaling form as the longest relaxation time in the relaxed state.
The solution 
reads~\cite{frank}
\begin{equation}
 {\tilde \phi}({\tilde r}, {\tilde t})= 
c_0 {\tilde R}({\tilde t})^{-3} 
\left[ 1-\{{\tilde r}/{\tilde R}({\tilde t})\}^2\right]^{1/m}
\label{sol.nonlFP}
\end{equation}
where $c_0= \Gamma[5/2+1/m]/(\pi^{3/2} \Gamma[1/m+1])$ with the Gamma
function, $\Gamma$, is the normalization factor and ${\tilde R}$ corresponds to the overall chain size, which evolves according to
\begin{eqnarray}
{\tilde R}({\tilde t}) = {\tilde R}_0\left[1+{\tilde t}/{\tilde \tau}_{ini}^{(0)} \right]^{\alpha_0}
\label{dynamics_exact}
\end{eqnarray}
with the exponent $\alpha_0=(3\nu-1)/(9\nu-2)$ and the time constant
${\tilde \tau}_{ini}^{(0)}=\alpha_0 m {\tilde R}_0^{1/\alpha_0}/(2
c_0^m)$.
Plots of the solution are shown in Fig.\ref{fig.fp}b along with the results of numerical calculation carried out using the initial condition of a step-like distribution.
In the strict sense, solution (\ref{sol.nonlFP}) is valid only for the initial condition $ {\tilde \phi}_0= c_0 {\tilde R}_0^{-3} [ 1-({\tilde r}/{\tilde R}_0)^2]^{1/m}$. 
Nevertheless, the figure shows that the dependence on the initial distribution is not significant.
The analytical results correspond well with those of the numerical calculation, particularly for the mean size of a chain ${\tilde R}_m \equiv \int \tilde{r} {\tilde \phi}(\tilde{r}) d  {\tilde {\mathbf r}}/\int {\tilde \phi}(\tilde{r}) d  {\tilde {\mathbf r}} \ (\simeq {\tilde R})$ (Fig.\ref{fig.fp}a).

\paragraph{Uniform Approximation}
Except for fortunate cases as above, one has to rely on the numerical
computation. In such cases, it would be desirable to find out an
approximation scheme allowing for a physical insight into the process.
In this spirit, we propose {\it uniform approximation}, in which the spatial distribution of monomers are regarded as uniform.
Note that almost the same attitude was adopted to study the stretching dynamics of a tethered polymer (monoblock picture)~\cite{Brochard}.

Let us illustrate it by applying to expansion dynamics.
In the expansion process, monomers are assumed to be uniformly confined in a spherical cavity of size $R(t)$, which is the size of the polymer chain at the moment(Fig.~\ref{fig1}). The volume fraction $\phi(t) \simeq a^3 N_0 /R(t)^3$ is connected with the blob size $\xi(t) \simeq a \phi(t)^{\nu/(1-3\nu)}$. Then, the excess free energy of the confinement is written as
\begin{eqnarray}
\Delta F/(k_BT) \simeq R(t)^3/\xi(t)^3 \simeq [a/R(t)]^{3/(3\nu-1)}N_0^{3\nu/(3\nu-1)}.
\label{F}
\end{eqnarray}
A nonlinear dependence on the chain length: $\Delta F \sim N_0^{9/4}$ for a chain with $\nu=3/5$ (in good solvent) and $\Delta F \sim N_0^{3}$ with $\nu=1/2$ (in $\theta$ solvent) is a characteristic of the SC regime~\cite{Sakaue_EPL, Sakaue_Macro2}. 
By noting the velocity gradient $\nabla v  \simeq {\dot \xi}(t)/\xi(t)$,
the overall dissipation is evaluated as
\begin{eqnarray}
T{\dot S(t)}  \simeq \eta R(t) {\dot R(t)}^2.
\end{eqnarray}
Balancing the rate of the free energy change and the dissipation, we obtain the following solution
\begin{eqnarray}
\dot{\Delta  F(t)} = -T {\dot S(t)} \Leftrightarrow R(t) = R_0\left(1+t/\tau_{ini} \right)^{\alpha}
\label{dynamics_1}
\end{eqnarray}
with the exponent $\alpha=(3\nu-1)/(9\nu)$ and a time scale
$\tau_{ini} = \tau_0 ( R_0/aN_0^{1/3})^{1/\alpha}$.
Again, this can be rescaled with the characteristic time $\tau_{ex}$ so
that we obtain the same form as in eq. (\ref{dynamics_exact}); ${\tilde
R(t)} =R(t)/R_F = {\tilde R_0}(1 +{\tilde t}/{\tilde \tau}_{ini})^{\alpha}$
with the rescaled time ${\tilde \tau}_{ini} =\tau_{ini} / \tau_{ex} = ({\tilde R}_0)^{1/\alpha}$.
This time ${\tilde \tau}_{ini}$ designates an initial transient period, which depends on the degree of the initial compression $C_{com} \equiv \phi_0/\phi_F$ as ${\tilde \tau}_{ini} = C_{com}^{3\nu/(1-3\nu)}$.
At ${\tilde t} > {\tilde \tau}_{ini}$, a sub-diffusion like scaling regime ${\tilde R(t)} \simeq {\tilde t}^{\alpha}$ follows.
All these results agree reasonably well with the exact solution except for a slight difference in the dynamic
exponent. 
In particular, the uniform approximation is able to predict the correct
characteristic time $\tau_{ex}$ of expansion.
This success traces back to the correct evaluation of the driving and
dissipation mechanisms based on the fundamental
length scale $\xi$.

\paragraph{Ejection}
Consider a long polymer  initially confined in a spherical cavity of size $R_0$.
Through a small hole (with a size comparable to that of monomers), the
polymer escapes (Fig.~\ref{fig1}).
This process of polymer ejection, i.e., a polymeric version of effusion,
 has recently attracted considerable attention inspired by the
 biological problem such as a DNA packed into and unpacked out of a phage capsid. From a fundamental point of view, there have also been many theoretical and numerical attempts to understand how a flexible chain is ejected from a pore~\cite{Sung, Muthukumar, Luijten, Yoemans}. Even in this simplest setup, however, the situation is yet controversial. It was reported that the Monte Carlo estimate of the ejection time $\tau_{ej}$ can be fit well by the formula
\begin{eqnarray}
\tau_{ej}^{(D)} \sim N_0 \left(N_0/\phi_0\right)^{1/3\nu}
\label{tau_dject_D}
\end{eqnarray}
which may be obtained by assuming $\Delta F \simeq \Delta F_{ideal} \sim N_0$~\cite{Muthukumar}. One of the latest works criticized eq.~(\ref{tau_dject_D}) by emphasizing the importance of the notion of SC, and proposed the following formula based on the systematic Monte Carlo simulation~\cite{Luijten}:
\begin{eqnarray}
\tau_{ej}^{(m)} \sim N_0^{1+\nu}/\Delta \mu \sim N_0^{1+\nu}\phi_0^{1/(1-3\nu)}
\label{app_scaling}
\end{eqnarray}
where $\Delta \mu = \Delta F/N_0$ is the free energy of the confinement
(eq.~(\ref{F})) per monomer, i.e., the chemical potential difference as
a driving force. However, the origin of this scaling is unclear. Indeed,
it was proposed as a translocation time of a polymer driven by strong
and constant $\Delta \mu$~\cite{Kantor}, which would be asymptotically
valid in the limit of the strong driving force~\cite{Sakaue_PRE}. In the
present case of the ejection, $\Delta \mu$ is far below such an
asymptote and gradually decreases with the process advanced. Moreover,
eq.~(\ref{app_scaling}) shows an unexpected opposite trend of the
chain-length dependence for polymers in good and $\theta$ solvent:
$\tau_{ej}^{(m)} \sim N^{7/20}$ for $\nu=3/5$ and $\tau_{ej}^{(m)} \sim
N^{-1/2}$ for $\nu=1/2$. 
In such a situation, it is pertinent to go beyond the scaling formula and to clarify the underlying physics.

In principle, one can numerically solve eq.~(\ref{eq_solve}) with the boundary condition ensuring a vanishing flux at the wall except for the pore region. Here, we shall apply the method of uniform approximation to gain useful insight.
The free energy is essentially the same as that in eq.~(\ref{F}), but now the dynamical variable is the number of monomers $N(t)$ left inside the cavity, while the cavity size $R_0$ remains constant: $\Delta F/(k_BT) \simeq (a/R_0)^{3/(3\nu-1)}N(t)^{3\nu/(3\nu-1)}$.
At a particular moment during the ejection, only the monomer residing at
the pore feels a resultant osmotic driving force $\sim k_B T/\xi(t)$. This affects the other monomers yet interior cavity, which would be quantified by a distance dependent response function. A linear response theory then connects the response to the correlation function of the concentration fluctuation. This consideration indicates that the overall dissipation is dominated by the proximity of the pore within the range of the correlation length $\xi(t)$, where there exists a velocity gradient of segments of the order $\sim a {\dot N(t)}/\xi(t)$. Therefore, the dissipation term is evaluated as
$T {\dot S(t)} \simeq \eta \left[ {\dot N(t)} a /\xi(t)\right]^2 \xi(t)^3 $.
The time evolution is obtained as
\begin{eqnarray}
\dot{\Delta F(t)} = -T {\dot S(t)} \Leftrightarrow N(t)=N_0\left(1+t/\tau_1 \right)^{\beta}
\label{driven_process}
\end{eqnarray}
with the exponent $\beta = (1-3\nu)/(2(1-\nu))$ and time constant $\tau_1 \simeq \tau_0 \phi_0^{(1+\nu)/(1-3\nu)}N_0$.
The dynamic feature in eq.~(\ref{driven_process}) can be well compared with the recent result of the stochastic rotation dynamic simulation~\cite{Yoemans}.
This osmotic driven process ceases at $\tau_2$ that is determined by
$N(\tau_2) \simeq  g_0 \Leftrightarrow \tau_{1,2} \equiv \tau_1 + \tau_2
= \tau_0 \phi_0^{-(\nu+2)/(3\nu)}N_0^{(2+\nu)/(3\nu)}$, in which
$g_0\simeq (R_0/a)^{1/\nu}$ is the number of monomers corresponding to the relaxed chain size $R_0$. After $\tau_2$, there is no driving force, and the chain diffusively escapes from the cavity, which takes a time $\tau_D \simeq R_0^2/D_{N_0}$ with the diffusion coefficient $D_{N_0}$ of the polymer. Thus, we arrive at the formula for the total ejection time $\tau_{ej} =\tau_2 + \tau_{D} \simeq \tau_{1,2}+ \tau_D$, where the second near-equality holds in the situation of our interest, i.e., $R_F \gg R_0$. 

Let us see how this can be compared with previous simulations. We note that  $\tau_{ej}$ was so far measured in a model system in which the effect of the induced flow of solvent, i.e., hydrodynamic interactions, was neglected. To cope with this situation, we need to modify the dissipation term as
$T {\dot S(t)} \simeq \eta a^3 [{\dot N(t)}]^2 g(t) $.
This modification changes the dynamic exponent to $\beta = (1-3\nu)/(3(1-\nu))$ and the time constant to $\tau_1 = \tau_0 \phi_0^{2/(1-3\nu)}N_0$ and $\tau_{1,2} = \tau_0 \phi_0^{-1/\nu}N_0^{1/\nu}$. 
Identifying the diffusion coefficient as $D_{N_0} \simeq k_B T/(\eta a N_0)$, the diffusion time is evaluated as $\tau_D = \tau_0 \phi_0^{-2/3}N_0^{5/3}$. 
In the limit of strong compression $R_0 \ll R_F$ with large $N_0$ and $R_0$, $\tau_{ej}$ asymptotically approaches the simple scaling form $\simeq \tau_0 (R_0/a)^{3/\nu}$ with no dependence on $N_0$. However, in most practical situations, the two contributions $\tau_{1,2}$ and $\tau_D$ are mixed. 
\begin{figure}[h]
\includegraphics[width=0.36\textwidth]{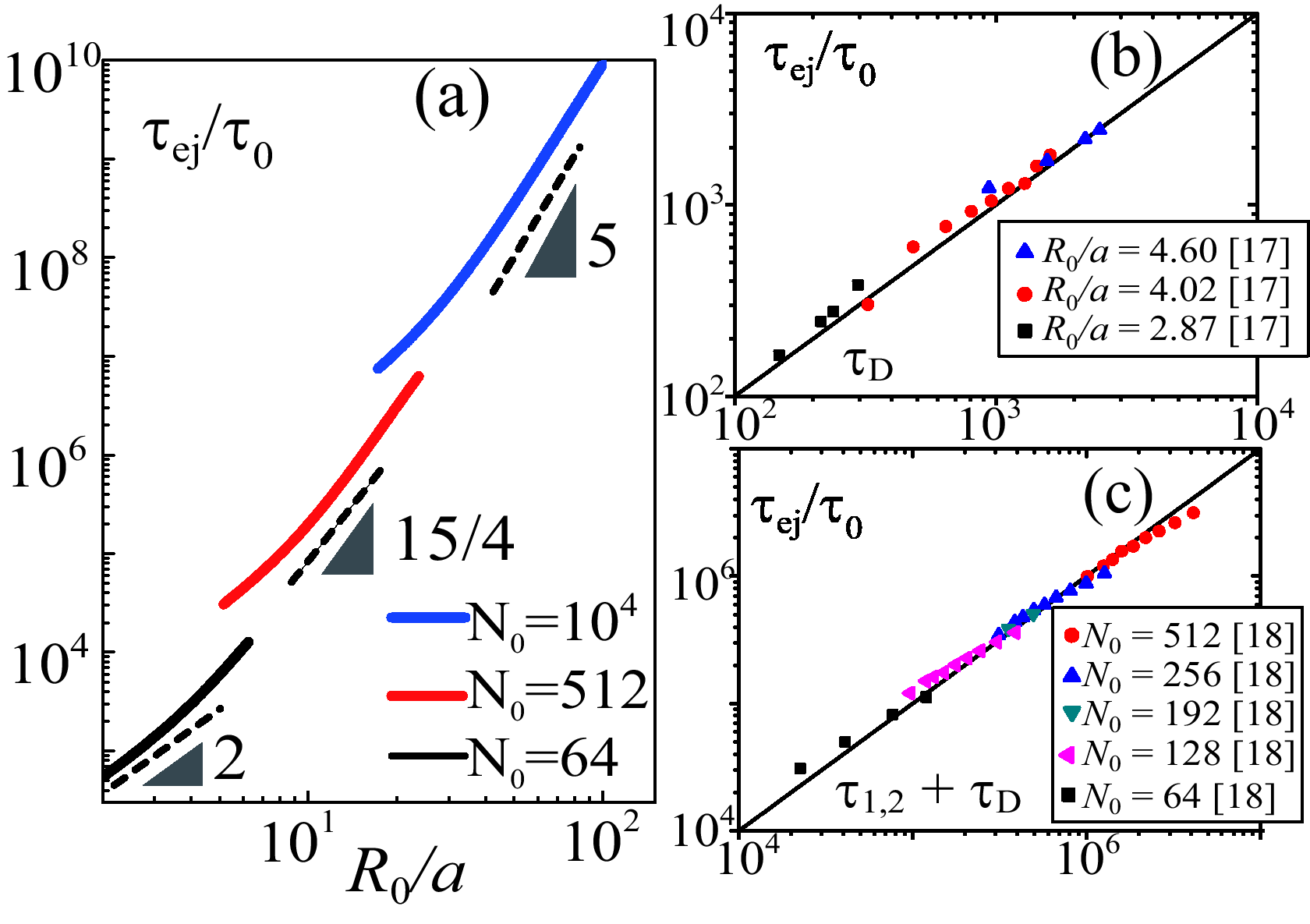}
\caption{Comparison of predicted ejection
 time with that obtained from previous numerical experiments. (a) Double logarithm plot of
 $\tau_{ej}$ as a function of $R_0$ for various system sizes. (b)
 Scaling plot of $\tau_{ej}$ vs. $\tau_D$ for small systems. (c)
 Scaling plot of $\tau_{ej}$ vs. $\tau_{1,2} + \tau_D$ for
 intermediate systems. Numerical data of $\tau_{ej}$ in (b) and (c) were extracted from~\cite{Muthukumar} and~\cite{Luijten}, respectively.}
\label{fig3}
\end{figure}
Analyzing various reported data in terms of our formula strongly indicates that the result in ref.~\cite{Muthukumar} (with small $N_0$
and $R_0$) corresponds to the late-stage diffusion dominated regime. 
In contrast, ref.~\cite{Luijten} employing larger $N_0$ and $R_0$ most probably monitors the mixing of the driven-ejection and diffusion regimes.
The asymptotic regime is realized only for extremely long chains.
Examples of the analysis (with $\nu=3/5$) are shown in Fig.~\ref{fig3}, in which the dependence of $\tau_{ej}$ on $R_0$ shows a clear crossover from the diffusion dominated regime ($\tau_{ej} \simeq \tau_D \sim R_0^2$, whose scaling form is close to eq.~(\ref{tau_dject_D})) to the asymptotic ($\tau_{ej} \simeq \tau_{1,2} \sim R_0^{5}$), between which there is a region where eq.~(\ref{app_scaling}): $\tau_{ej} \sim R_0^{15/4}$ looks valid. 
In addition, scaling plots of reported ejection times provide more
concrete evidence that ref.~\cite{Muthukumar} and ref.~\cite{Luijten},
respectively, monitor the ejection processes in the small and
intermediate size systems in accordance with the above scenario. 
We obtain very good data collapse without any fitting parameters
(Fig. 3b and c) \cite{note1}.
This supports the suitability of our analysis. 

In conclusion, we have proposed a general framework for the decompression dynamics of polymers. Aside from a basic equation of the nonlinear diffusion type, uniform approximation has been shown to be useful. We applied the idea and method to the problems of polymer expansion, unfolding and the ejection. We believe that, for the ejection problem in particular, this provides the first-ever consistent description with both the driving force and the dissipation mechanism clearly identified.

We would like to end with some remarks and perspectives;
(I) One may wonder how the ejection process of a polymer from a planar
slit numerically studied in \cite{Luijten,panja:2008b} is described. We point out the softness of the two-dimensionally confined polymer, i.e., its elastic modulus $\sim (k_B T)/R_{F_2}^3$ is small compared to the driving osmotic force $\sim (k_BT)/R_0$, where $R_{F_2} \sim N_0^{3/4}$ and $R_0$ are the lateral size of the confined chain and the slit distance. In contrast, in the ejection process from a closed cavity, both the elastic modulus and the driving force are dictated by the mesh size $\xi$. In the former case, the entire chain cannot respond to the ejection force acting on a particular monomer residing at the pore immediately, and the highly nonequilibrium process of the front evolution associated with the tension propagation along the chain would occur~\cite{Sakaue_PRE}, while in the latter, the pressure drop is transmitted not along the chain, but radially from the pore. These differences in the underlying physics are indeed a consequence of the different confinement regimes~\cite{Sakaue_EPL, Sakaue_Macro2}.
(II) The current description based on the analogy with the semidilute solution cannot be applied to the very dense system with $\phi \simeq 1$. In such a densely compressed situation, one has to deal with solvent diffusion and permeation into the globular chain along with the decompression. In addition, a self-entanglement may be formed for a very long chain, which alters the dynamics qualitatively~\cite{Arrest1,Arrest2,Yoshinaga}.
(III) We have assumed that the polymer is intrinsically flexible. As discussed in ref.~\cite{Sakaue_Macro2}, the effect of the chain stiffness would have a pronounced effect in a confined space, which alters the scaling structure of the free energy. 
We hope that these points would be clarified in future studies.

TS acknowledges the support from Grant-in-Aid for Young Scientists (No. 20840027).
NY acknowledges the support from the JSPS (No. 7662).
Part of the numerical calculations were
carried out on Altix3700 BX2 at YITP at Kyoto University.


\begin{thebibliography}{50}
\bibitem{note1} $\tau_0$ 
	depending on details of simulations is used merely
	for making the proportional factor one. We have used
	$\tau_0 = 7.2$ Monte Carlo steps for \cite{Luijten} and $\tau_0 = 26.4$ for \cite{Muthukumar}.
\bibitem{note2} The value is modified as $m=1/(3 \nu -1)$ without
	hydrodynamics since the transport coefficient is  $L
	\simeq 1/(\eta a)$.
\end{thebibliography}
\end{document}